\documentclass[
    ,final            
  ]
  {aipproc}

\layoutstyle{6x9}

\newcommand{\NP}[1]{{ Nucl.\ Phys.\ } {\bf  #1}}

\newcommand{\PL}[1]{{ Phys.\ Lett.\ } {\bf  #1}}

\newcommand{\PRL}[1]{{ Phys.\ Rev.\ Lett.\ } {\bf  #1}}

\newcommand{\lsim}{\raise.3ex\hbox{$<$\kern-.75em\lower1ex\hbox{$\sim$}}}
\newcommand{\ima}{{\mbox{Im}\,}}
\newcommand{\rea}{{\mbox{Re}\,}}
\newcommand{\be}{\begin{equation}}
\newcommand{\ee}{\end{equation}}

\begin{document}

\title{Large $N_c$ behavior of light resonances
in meson-meson scattering
\footnote{To appear in the proceedings of the HADRON03
conference, Aschaffenburg, Germany. September 2003}
}
\author{J.R.Pel\'aez}{address={Departamento de F\'{\i}sica Te\'orica II, 
  Universidad Complutense, 28040 Madrid, Spain}
}

\begin{abstract}
By scaling the parameters of  meson-meson
unitarized Chiral Perturbation Theory amplitudes
according to the QCD large $N_c$ rules, 
one can  study the spectroscopic nature of light
meson resonances. The scalars $\sigma$, $\kappa$
$f_0(980)$ and, possibly, the  $a_0(980)$ do not seem to behave as $\bar qq$ states,
in contrast to the vectors $\rho(770)$ and $K^*(892)$.
The behavior shown by the scalars is naturally explained
in terms of diagrams with intermediate $\bar q\bar q qq$-like states.
Here we review our recent study and 
show how the results do not depend on the different fits
to data.
\end{abstract}

\maketitle
\section{ Introduction}
Although QCD is 
firmly established as the theory of
strong interactions it becomes non-perturbative at low energies, 
and gives only little help to address the existence and nature
of the lightest scalar mesons. Alternatively
Chiral Perturbation Theory (ChPT) \cite{chpt1}
has been devised as the QCD low energy Effective Lagrangian
built as the most general derivative expansion
respecting its symmetries, 
containing only $\pi, K$ and $\eta$ mesons. These particles 
are the QCD low energy degrees of freedom
since they are Goldstone bosons of the QCD spontaneous
chiral symmetry breaking.
For meson-meson scattering  ChPT is an expansion in even
powers of momenta, $O(p^2), O(p^4)$...,
 over a scale $\Lambda_\chi\sim4\pi f_0\simeq 1\,$GeV.
Since the $u$, $d$ and $s$ quark
masses are so small compared with $\Lambda_\chi$ they
are introduced as perturbations, giving rise to the 
$\pi, K$ and $\eta$ masses, counted as $O(p^2)$. 
At each order, ChPT is the sum of {\it all terms}
 compatible with the symmetries,
multiplied by ``chiral'' parameters, that absorb
loop divergences order by order, yielding finite results.
The leading order is universal since there is only one parameter,  $f_0$, 
that sets the scale of spontaneous
chiral symmetry breaking.
Different underlying dynamics manifest themselves with different 
higher order  parameters.
In Table I are listed the parameters that determine
meson-meson scattering up to $O(p^4)$, called $L_i$.
As usual after renormalization, they
depend on an arbitrary 
 regularization scale, as 
$
 L_i(\mu_2)=L_i(\mu_1)+\Gamma_i\log(\mu_1/\mu_2)/{16\pi^2},
$
where $\Gamma_i$ are constants given in \cite{chpt1}.
In physical observables the $\mu$ dependence is canceled
with that of the loop integrals. 
 
The large $N_c$ expansion \cite{'tHooft:1973jz}
is the only
analytic approximation to QCD in the whole
energy region, also
providing a clear definition of $\bar qq$ states, that become
bound states when $N_c\rightarrow\infty$.
The $N_c$ scaling
of the $L_i$ parameters has been given in
\cite{chpt1,chptlargen}, and is listed in Table I. 
In addition, the $\pi,K,\eta$ masses scale as $O(1)$ and 
$f_0$ as $O(\sqrt{N_c})$.
However, it is not known at what scale $\mu$ to apply the large $N_c$
scaling, and it has been pointed out that the logarithmic terms 
can be rather large for $N_c=3$ \cite{Pich:2002xy}.
The scale dependence is certainly
suppressed by $1/N_c$ for $L_i=L_2,L_3,L_5,L_8$, but not for
$2L_1-L_2, L_4,L_6$ and $L_7$. 
Customarily, the uncertainty in the $\mu$ 
where the $N_c$ scaling applies is estimated 
varying $\mu$ between 0.5 and 1 GeV \cite{chpt1}.
We will check that this estimate is correct with the vector mesons,
firmly established as $\bar qq$ states.

\begin{table}[hbpt]
\begin{tabular}{|c||c||c||c|c|c|}
\hline
  \tablehead{1}{c}{b}{$O(p^4)$ \\ Parameter} &
  \tablehead{1}{c}{b}{ 
$N_c$\\scaling} &
  \tablehead{1}{c}{b}{\hspace*{0.6cm}ChPT\hspace*{0.6cm}\\$\mu=770\,$MeV } &
  \tablehead{1}{c}{b}{\hspace*{0.6cm}IAM I\hspace*{0.6cm}\\$\mu=770\,$MeV }&  
  \tablehead{1}{c}{b}{\hspace*{0.6cm}IAM II\hspace*{0.6cm}\\$\mu=770\,$MeV } &
  \tablehead{1}{c}{b}{\hspace*{0.6cm}IAM III\hspace*{0.6cm}\\$\mu=770\,$MeV } 
\\
\hline
$L_1$
& $O(N_c)$
& $0.4\pm0.3$
& $0.56\pm0.10$ 
& $0.59\pm0.08$
& $0.60\pm0.09$
\\
$L_2$
& $O(N_c)$
& $1.35\pm0.3$ 
& $1.21\pm0.10$ 
& $1.18\pm0.10$
& $1.22\pm0.08$\\
$L_3 $  &
 $O(N_c)$&
 $-3.5\pm1.1$&
$-2.79\pm0.14$ 
&$-2.93\pm0.10$
& $-3.02\pm0.06$
\\
$L_4$
& $O(1)$
& $-0.3\pm0.5$& $-0.36\pm0.17$ 
& $0.2\pm0.004$
& 0 (fixed)\\
$L_5$
& $O(N_c)$
& $1.4\pm0.5$& $1.4\pm0.5$ 
& $1.8\pm0.08$
& $1.9\pm0.03$
\\
$L_6$
& $O(1)$
& $-0.2\pm0.3$& $0.07\pm0.08$ 
&$0\pm0.5$
&$-0.07\pm0.20$\\
$L_7 $  & $O(1)$ & 
$-0.4\pm0.2$&
$-0.44\pm0.15$ &
$-0.12\pm0.16$&
$-0.25\pm0.18$
\\
$L_8$
& $O(N_c)$
& $0.9\pm0.3$& $0.78\pm0.18$ 
&$0.78\pm0.7$
&$0.84\pm0.23$\\
\hline
$2L_1-L_2$
& $O(1)$
& $-0.55\pm0.7$& $0.09\pm0.10$ 
&$0.0\pm0.1$
&$-0.02\pm0.10$\\
\hline
\end{tabular}
\caption{$O(p^4)$ chiral parameters ($\times10^{3}$) and their $N_c$ scaling.
In the ChPT column, $L_1,L_2,L_3$ come from
\cite{BijnensGasser} and  the rest from  \cite{chpt1}. The 
IAM
columns correspond to different fits \cite{GomezNicola:2001as}} 
\label{eleschpt}
\end{table}

ChPT is a low energy expansion, but in recent
years it has been extended to higher energies by means of unitarization 
\cite{GomezNicola:2001as,Dobado:1996ps,Oller:1997ng,Guerrero:1998ei,Oller:1997ti}. 
The main idea is that when projected into partial waves of definite
angular momentum $J$ and isospin $I$, physical amplitudes $t$ should satisfy
an elastic unitarity condition:
\begin{equation}
  \ima t =\sigma \vert t\vert^2 \quad\Rightarrow \quad\ima \frac{1}{t}=-\sigma \quad\Rightarrow\quad
t=\frac{1}{\rea t^{-1} - i \sigma},
\end{equation}
where $\sigma$ is the phase space of the two mesons, a well known function.
However, from the right hand side we note that to have a unitary
amplitude we only need $\rea t^{-1}$, and for that 
we can use the ChPT expansion; this is the
Inverse Amplitude Method (IAM) \cite{Dobado:1996ps}.  
The IAM generates the $\rho$, $K^ *$, $\sigma$ and $\kappa$ resonances 
not initially present in ChPT, ensures unitarity
in the elastic region and respects the ChPT expansion.
When inelastic two-meson processes are present the IAM generalizes 
to $T\simeq(\rea T^{-1}-i \Sigma)^{-1}$ where 
$T$ is a matrix containing all partial waves 
between all physically accessible states whereas
$\Sigma$ is a diagonal
matrix with their phase spaces,
again well known 
\cite{GomezNicola:2001as,Oller:1997ng,Guerrero:1998ei,Oller:1997ti}. 
With this generalization it was recently
shown  \cite{GomezNicola:2001as} that, using the one-loop ChPT calculations,
the IAM generates the $\rho$, $K^ *$, $\sigma$, $\kappa$,
$a_0(980)$,
$f_0(980)$ and the octet $\phi$, describing two body $\pi$, K or $\eta$
scattering up to 1.2 GeV. Furthermore, it has
the correct low energy expansion, with chiral parameters
compatible with standard ChPT, shown in Table I. 
Different IAM fits \cite{GomezNicola:2001as} 
are due to different ChPT truncation schemes
equivalent up to $O(p^ 4)$ and the 
estimates of the data systematic error.  

Those IAM results have been recently 
used \cite{Pelaez:2003dy} to study the large $N_c$ 
behavior of the scattering amplitudes and the poles associated to resonances.
The large $N_c$ results are similar for all IAM sets, and to illustrate it,
we show here the results of set III, whereas in \cite{Pelaez:2003dy}
we used set II, reaching the same conclusions. 
Note that these ChPT amplitudes are fully
renormalized,
and therefore scale independent. Hence
all the QCD $N_c$ dependence appears correctly
through the $L_i$ 
and cannot  hide in any spurious parameter.

\section{Results}

Let us then scale $f_0\rightarrow f_0 \sqrt{N_c/3}$
and $L_i(\mu)\rightarrow L_i(\mu)(N_c/3)$ for $i=2,3,5,8$, keeping
the masses and $2L_1-L_2,L_4,L_6$ and $L_7$ constant.
In Fig.1 we show, for increasing $N_c$, the modulus of the 
$(I,J)=(1,1)$ and $(1/2,1)$ amplitudes
with the Breit-Wigner shape of the $\rho$
and $K^*(892)$ vector resonances, respectively.
There is always a peak at an almost constant position,
becoming 
narrower as $N_c$ increases.
We also show the evolution
of the $\rho$ and $K^*$ pole positions,
related to their mass and width as $\sqrt{s_{pole}}\simeq M-i \Gamma/2$.
We have normalized both $M$ and $\Gamma$
to their value at $N_c=3$ in order to compare
with the $\bar{q} q$ expected behavior:
$M_{N_c}/M_3$ constant and  $\Gamma_{N_c}/\Gamma_3\sim 1/N_c$
The agreement is remarkable within
the gray band that covers the uncertainty 
$\mu=0.5-1\,$GeV where to apply the large $N_c$ scaling.
We have checked that outside that band, the behavior starts 
deviating from that of $\bar qq$ states, which confirms that
the expected scale range where the large $N_c$ scaling 
applies is correct.
\begin{figure}[hbpt]
\includegraphics[width=0.67\textheight]{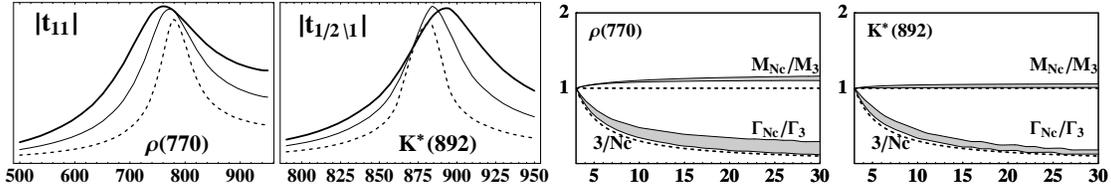}
\caption
{Left: Modulus 
of $\pi\pi$ and $\pi K$ elastic amplitudes versus $\sqrt{s}$
for $(I,J)=(1,1),(1/2,1)$:  
$N_c=3$ (thick line), $N_c=5$ (thin line) and $N_c=10$ 
(dotted line), scaled at $\mu=770\,$MeV.
Right: $\rho(770)$ and $K^*(892)$ pole positions: $\sqrt{s_{pole}}\equiv M-i\Gamma/2$
versus $N_c$. The gray areas cover the uncertainty $N_c=0.5-1\,$GeV. The dotted lines
show the expected $\bar q q$ large $N_c$ scaling.}
\label{fig:f2}
\end{figure}

In Fig.2, in contrast, all over the $\sigma$ and $\kappa$ regions
the $(0,0)$ and $(1/2,0)$ amplitudes decrease as $N_c\rightarrow\infty$.
Their associated poles 
show a totally
different behavior, since 
{\em their width grows with} $N_c$, in
conflict with a ${\bar qq}$ interpretation. 
(We keep the $M$, $\Gamma$ notation, but now as definitions).
This is also suggested using the ChPT leading order
unitarized amplitudes with a 
regularization scale \cite{Oller:1997ti,Harada:2003em}.
In order to determine their spectroscopic nature,
we note that {\em in the whole} $\sigma$ and $\kappa$ regions,
$\ima t\sim O(1/N_c^2)$ and $\rea t\sim O(1/N_c)$.
Imaginary parts are generated from s-channel intermediate 
physical states. If it was a  
$\bar qq$ meson, with mass $M\sim O(1)$ and $\Gamma\sim 1/N_c$, 
we would expect $\ima t\sim O(1)$ and a peak
at $\sqrt{s}\simeq M$, as it is indeed
the case of the $\rho$ and $K^*$. Therefore, 
from $\bar q q$ states, the $\sigma$ and $\kappa$  
can only get real contributions from $\rho$ or $K^*$ t-channel exchange, 
respectively. 
The leading s-channel contribution for the $\kappa$ 
comes from $\bar q \bar q qq$ (or two
meson) states, which  are predicted to unbound and  
become the meson-meson continuum when $N_c\rightarrow\infty$ 
\cite{Jaffe}. The same interpretation holds for the sigma,
but in the large $N_c$ limit $\bar q \bar q qq$ 
and glueball exchange count the same. Given the fact that
glueballs are expected to have masses above 1 GeV, and that
the $\kappa$ is a natural $SU(3)$ partner of the $\sigma$,
a dominant $\bar q \bar q qq$  component for the $\sigma$ seems the
most natural interpretation, although it could certainly have some glueball
mixing.
\begin{figure}[hbpt]
\includegraphics[width=0.67\textheight]{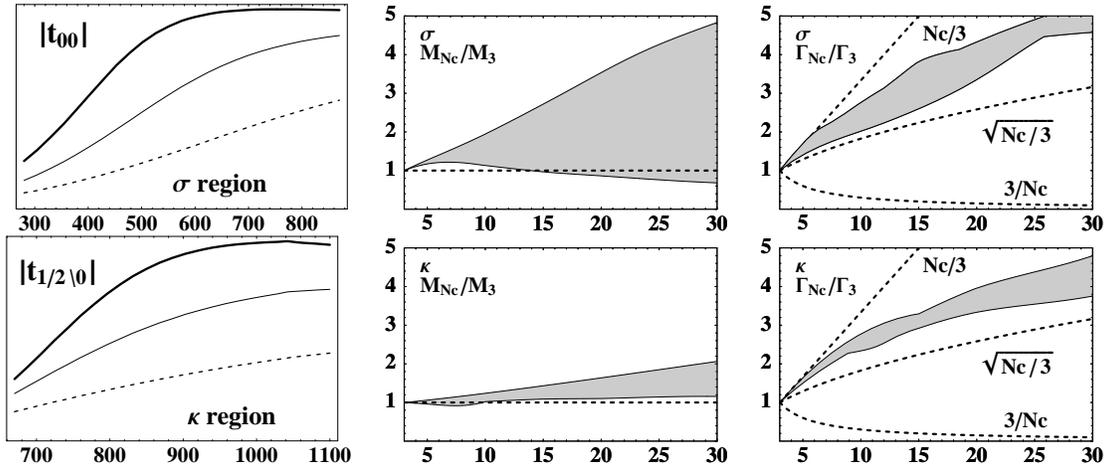}
\caption
{Top) Right:
 Modulus of the $(I,J)=(0,0)$ scattering amplitude, versus $\sqrt{s}$
for $N_c=3$ (thick line), $N_c=5$ (thin line) and $N_c=10$ 
(dotted line), scaled at $\mu=770\,$MeV. Center: $N_c$ evolution
of the $\sigma$ mass. Left: $N_c$ evolution
of the $\sigma$ width. Bottom: The same but for the $(1/2,0)$ amplitude
and the $\kappa$.
}
\label{fig:f2}
\end{figure}

The large $N_c$ behavior of the $(0,0)$ amplitude
in the vicinity of the $f_0(980)$ is shown in Fig.3.
This resonance and the $a_0(980)$ 
are more complicated due to the distortions 
caused by the nearby $\bar KK$ threshold.
We see that the characteristic sharp dip of the $f_0(980)$ 
vanishes when $N_c\rightarrow\infty$, at variance with
a $\bar qq$ state. For $N_c>5$ it follows again the $1/N_c^2$ scaling compatible 
with $\bar q\bar q qq$ states or glueballs.
The $a_0(980)$ behavior, shown in Fig.4, is more complicated.
When we apply the large $N_c$ scaling
at $\mu=0.55-1\,$ GeV, its
 peak disappears, suggesting that this is not a $\bar q q$
state, and $\ima t_{10}$
follows roughly the $1/N_c^2$ behavior
in the whole region \footnote{ The idea of this work
and the pole movements
were presented by the author in two workshops \cite{Pelaez:2003dy}.
While completing the calculations and the manuscript
the results without the scale uncertainties
have been confirmed \cite{Uehara:2003ax}  for all resonances, using the
approximated IAM \cite{Oller:1997ng}.}. 
However, as shown in Fig.5, the peak does not vanish at
large $N_c$ if we take $\mu=0.5\,$GeV. Thus we cannot rule
out a possible $\bar q q$ nature, or a sizable mixing,
although it shows up in
an extreme corner of our uncertainty band. 
For other recent large $N_c$ arguments in a chiral context
see \cite{Cirigliano:2003yq}.
\begin{figure}[hbpt]
\includegraphics[width=0.5\textheight]{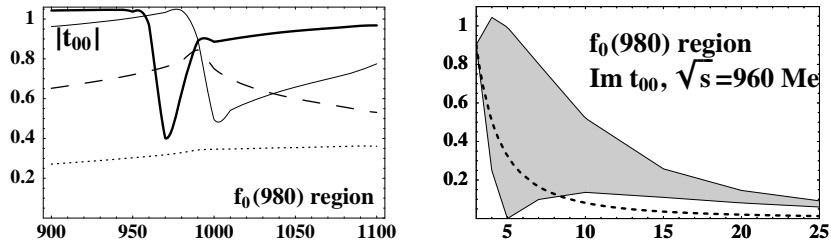}
\caption
{Right:
 Modulus of a $(I,J)=(0,0)$ scattering amplitude, versus $\sqrt{s}$,
for $N_c=3$ (thick), $N_c=5$ (thin), $N_c=10$ 
(dashed) and $N_c=25$ 
(dotted), scaled at $\mu=770\,$MeV. Left: $\ima t_{00}$ 
versus $N_c$.}
\label{fig:f2}
\end{figure}
\begin{figure}[hbpt]
\includegraphics[width=0.67\textheight]{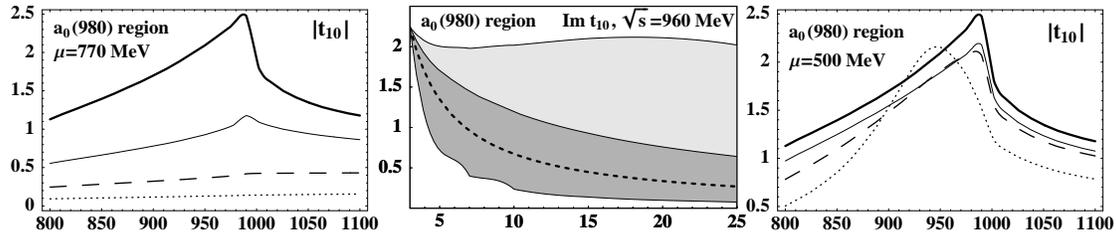}
\caption
{Right:
 Modulus of a $(I,J)=(1,0)$ scattering amplitude, versus $\sqrt{s}$,
for $N_c=3$ (thick), $N_c=5$ (thin), $N_c=10$ 
(dashed) and $N_c=25$ 
(dotted), scaled at $\mu=770\,$MeV. Right: scaled at $\mu=500\,$MeV.
Center: $\ima t_{00}$ versus $N_c$. The dark gray area covers 
the uncertainty $\mu=0.55-1\,$GeV, the light gray area from $\mu=0.5$ to $0.55\,$GeV.}
\label{fig:f2}
\end{figure}
\section{Conclusion}

We have shown that the QCD large $N_c$ scaling
of the unitarized meson-meson amplitudes
of  Chiral Perturbation Theory 
is in conflict with a $\bar qq$ nature for the lightest scalars
(not so conclusively for the $a_0(980)$), 
and strongly suggests a 
$\bar q \bar qqq$ or two meson main component, maybe with some mixing
with glueballs, when possible.

\begin{theacknowledgments}
I thank A. Andrianov, D. Espriu, A. G\'omez Nicola, F. Kleefeld,
R. Jaffe, E. Oset, J. Soto and  M. Uehara for their comments
and support from
the Spanish CICYT projects,
BFM2000-1326, BFM2002-01003 and the 
E.U. EURIDICE network contract no. HPRN-CT-2002-00311.
\end{theacknowledgments}

\bibliographystyle{aipproc}   

\begin{thebibliography}{99}

\footnotesize


\bibitem{chpt1}
S. Weinberg, Physica {\bf A96} (1979) 327.
J.~Gasser and H.~Leutwyler,
Annals Phys.\  {\bf 158} (1984) 142;
Nucl.\ Phys.\ B {\bf 250} (1985) 465.


\bibitem{BijnensGasser} J. Bijnens, G. Colangelo and J. Gasser,
\NP{B427} (1994) 427.


\bibitem{GomezNicola:2001as}
A.~G\'omez Nicola and J.~R.~Pel\'aez,
Phys.\ Rev.\ D {\bf 65} (2002) 054009 and
AIP Conf.\ Proc.\  {\bf 660} (2003) 102
[hep-ph/0301049].




\bibitem{'tHooft:1973jz}
G.~'t Hooft,
Nucl.\ Phys.\ B {\bf 72} (1974) 461.
E.~Witten,
Annals Phys.\  {\bf 128} (1980) 363.

\bibitem{chptlargen}
A.~A.~Andrianov,
Phys.\ Lett.\ B {\bf 157}, 425 (1985).
A.~A.~Andrianov and L.~Bonora,
Nucl.\ Phys.\ B {\bf 233}, 232 (1984).
D.~Espriu, E.~de Rafael and J.~Taron,
Nucl.\ Phys.\ B {\bf 345} (1990) 22
S.~Peris and E.~de Rafael,
Phys.\ Lett.\ B {\bf 348} (1995) 539

\bibitem{Pich:2002xy}
A.~Pich,
hep-ph/0205030.

\bibitem{Dobado:1996ps}
T.~N.~Truong,
Phys.\ Rev.\ Lett.\  {\bf 61} (1988) 2526.
\PRL{67}, (1991) 2260;
A. Dobado, M.J.Herrero and T.N. Truong, \PL{B235} (1990) 134.
A.~Dobado and J.~R.~Pelaez,
Phys.\ Rev.\ D {\bf 47} (1993) 4883.
Phys.\ Rev.\ D {\bf 56} (1997) 3057.

\bibitem{Oller:1997ng}
J.~A.~Oller, E.~Oset and J.~R.~Pelaez,
Phys.\ Rev.\ Lett.\  {\bf 80} (1998) 3452;
Phys.\ Rev.\ D {\bf 59} (1999) 074001
 and Phys.\ Rev.\ D {\bf 62} (2000) 114017.
M.~Uehara,
hep-ph/0204020.



\bibitem{Guerrero:1998ei}
F.~Guerrero and J.~A.~Oller,
Nucl.\ Phys.\ B {\bf 537} (1999) 459
[Erratum-ibid.\ B {\bf 602} (2001) 641].



\bibitem{Oller:1997ti}
J.~A.~Oller and E.~Oset,
Nucl.\ Phys.\ A {\bf 620} (1997) 438; 
Phys.\ Rev.\ D {\bf 60} (1999) 074023.

\bibitem{Pelaez:2003dy}
J.~R.~Pelaez,
hep-ph/0309292;
hep-ph/0307018 and
hep-ph/0306063.


\bibitem{Harada:2003em}
M.~Harada, F.~Sannino and J.~Schechter,
hep-ph/0309206.


\bibitem{Jaffe} R. L. Jaffe, Proceedings of the Intl. Symposium
on Lepton and Photon Interactions at High Energies. Physikalisches Institut, University of Bonn (1981) . ISBN: 3-9800625-0-3 

\bibitem{Cirigliano:2003yq}
V.~Cirigliano {\em et al.}
hep-ph/0305311.
N.~N.~Achasov,
hep-ph/0309118.
T.~Schaefer,
hep-ph/0309158.



\bibitem{Uehara:2003ax}
M.~Uehara,
hep-ph/0308241.



\end{thebibliography}

\end{document}